\documentclass[
aps,pre,
twocolumn,
groupedaddress,
amsmath,amssymb,
floatfix,
]{revtex4-2}

\usepackage{graphicx}  
\usepackage{dcolumn}  
\usepackage{bm}  

\begin{document}

\preprint{APS/123-QED}

\title{Modeling Ion-Specific Effects in Polyelectrolyte Brushes:\\
A Modified Poisson-Nernst-Planck Model}

\author{William J Ceely}  
 \email{william.ceely@cgu.edu}
\author{Marina Chugunova}  
 \email{marina.chugunova@cgu.edu}
\author{Ali Nadim}  
 \email{ali.nadim@cgu.edu}
\affiliation{
 Institute of Mathematical Sciences,
 Claremont Graduate University,
 Claremont, California, 91711, USA 
}

\author{James D Sterling}
 \email{jim\_sterling@kgi.edu}  
\affiliation{
 Henry E. Riggs School of Applied Life Sciences,
 Keck Graduate Institute,
 Claremont, California, 91711, USA 
}

\date{\today}

\begin{abstract}
Polyelectrolyte brushes consist of a set of charged linear macromolecules each
tethered at one end to a surface. An example is the glycocalyx which refers to
hair-like negatively charged sugar molecules that coat the outside membrane of all
cells. We consider the transport and equilibrium distribution of ions, and the
resulting electrical potential, when such a brush is immersed in a salt buffer
containing monovalent cations (sodium and/or potassium). The Gouy-Chapman model
for ion screening at a charged surface captures the effects of the Coulombic force
that drives ion electrophoresis and diffusion, but neglects non-Coulombic forces
and ion pairing. By including the distinct binding affinities of these counter-ions
with the brush, and their so-called Born radii, which account for Born forces
acting on them when the permittivity is non-uniform, we propose modified
Poisson-Nernst-Planck continuum models that show the distinct profiles that may
result depending on those ion-specific properties.
\end{abstract}

\maketitle


\section{Introduction}\label{intro}
Polyelectrolyte brushes composed of glycosaminoglycans (GAGs) serve as a model of
the glycocalyx that exists as an extracellular layer in all animals. A brush consists
of many strands that each have their own long, thin quasi-linear structure. The disaccharides
comprising the glycocalyx are different in each of the four mammalian GAGs including
heparan sulfate, chondroitin sulfate, keratin sulfate, and the only non-sulfated
GAG hyaluronic acid. The glycocalyx often serves as the first point of contact of
pathogens as they encounter endothelial and mucosal surfaces that contain GAGs as
well as tethered and secreted mucins. The glycocalyx may swell and shrink as a
polyelectrolyte brush in response to the surrounding water and salt. It is associated
with inflammation and immune system function. These are highly anionic brushes
that are neutralized by monovalent and divalent cations as well as histamine,
chemokines and cationic domains of proteins rich in arginine and lysine. A full
understanding of the structure and functions of the myriad of glycocalyx structures
can be elucidated through study of biophysical effects of ion-partitioning and
the nature of hydration and electrostatic effects in the glycocalyx. We therefore
further-simplify our models of the glycocalyx as a brush of uniform anionic fixed
charge density GAGs, with constant permittivity away from the interface, neutralized
only by monovalent ions potassium and sodium.

The traditional model for ion screening at a charged surface is the classical
diffuse double-layer. By this description, for a negatively charged glycosaminoglycan
(GAG) brush in a salt solution, the first layer is a negative surface charge,
while the second layer is composed of an excess of positively charged ions (such as Na$^{+}$
or K$^{+}$) and a deficit of negatively charged ions.
The cations are attracted towards the surface by the Coulombic
force that drives ion electrophoresis, but equally diffuse away from the surface
from zero net flux. Detailed descriptions of the double-layer can be found in many textbooks
\cite{bruus2007theoretical, kirby2010micro, lyklema1995fundamentals, russel1991colloidal},
however, the basic idea is that the steady state can be found using Poisson's
equation with the assumption that the unbound mobile cation(s) and anion(s) partition
according to a Boltzmann distribution. This is known as the Gouy-Chapman model and is 
described mathematically by the Poisson-Boltzmann equation \cite{fogolari2002poisson}.

The Gouy-Chapman model does not account for any non-Coulombic forces and also
ignores ion pairing between the brush and oppositely charged ions. Molecular
simulation data results \cite{sterling2021ion} show both varying permittivity and
ion pairing lead to important additional contributions of a
Born energy and a binding energy, respectively. In our previous
work \cite{ceely2023mathematical}, we introduced modified Poisson-Boltzmann (MPB)
models to incorporate both of these effects. Our models applied to a negatively
charged GAG brush in a salt solution show that the addition of ion pairing and
Born hydration forces can yield a positive Donnan potential (for certain brush
cation pairings), as well as result in a double-double-layer or an even more complex
structure in the vicinity of the brush-salt interface.

In \S \ref{deriv-model}, we present the mathematical model for the time-dependent
evolution of the system and consider the relevant physical effects that are present
in this new model in comparison with the steady state MPB model by utilizing
Poisson's equation and the continuity equation \cite{chen1992constant} via a
modified Nernst-Planck equation. This model is known
as a modified Poisson-Nernst-Planck (MPNP) model. In \S \ref{init_far}, we consider
initial conditions
that are far from equilibrium in order to highlight how each of these four terms
(diffusion, electrophoresis, Born force, and ion pairing)
evolve over time and play a role in the movement of ions en route to equilibrium.
We then consider a set of more physically realistic near-equilibrium
initial conditions in \S \ref{init_real} to study the
time evolution of a GAG/salt system. Finally, in \S \ref{two_cations}, we consider
a system with two
cations to study how binding reactions impact the transfer of cations.

While we derive our model through a conservation equation, previous works begin
with a free energy functional and derive a PB or PNP model via a variational
approach \cite{fogolari1997variational, li2009minimization, liu2018analysis, zhulina2011poisson}.
In \S \ref{energy-functional}, we develop the corresponding free energy functional
to our MPNP model to include
a binding energy density and Born solvation energy density and show through
numerical results that the steady state solution minimizes the total energy.

\section{A time-dependent model}\label{deriv-model}
The MPB models in \cite{ceely2023mathematical} focus
on the steady state solution of the glycosaminoglycan brush/salt system.
A complete mathematical description of the system should also include the
non-equilibrium dynamics prior to reaching steady state. To capture this transition,
we derive a time-dependent model in a domain, $\Omega$, with boundary, $\partial\Omega$
that incorporates ion pairing and Born energy due to varying permittivity.

As in \cite{ceely2023mathematical}, we start with Gauss's law, assume electrostatic
conditions, and that the total charge density is related to the concentrations of
the unbound ions, $\left[C_{i}\right]$ which gives us Poisson's equation for
electrostatics together with the boundary conditions,
\begin{subequations}
\begin{equation}
\bm{\nabla}\bm{\cdot}\left(-\varepsilon\bm{\nabla}\phi\right)
= \sum_{i} z_{i}N_{A}e\left[C_{i}\right], \qquad \text{in} \ \Omega,
\end{equation}
\begin{equation}
\bm{n}\bm{\cdot}\left(\varepsilon\bm{\nabla}\phi\right) = \sigma,
\qquad \text{on} \ \partial\Omega,
\end{equation}
\end{subequations}
where $\varepsilon$ is the permittivity, $z_{i}$ is the valence, $N_{A}$ is Avogadro's
number, $e$ is the charge of an electron, $\bm{n}$ is the outward unit normal vector
on $\partial\Omega$, and $\sigma$
is the surface charge density on the boundary. We also consider a modified
Nernst-Planck equation, which is the conservation equation for each of the
ion species undergoing diffusion, chemical reaction, and drift due to an external force,
\begin{subequations}
\begin{equation}
\frac{\partial \left[C_{i}\right]}{\partial t} + \bm{\nabla}\bm{\cdot}\bm{J}_{i} = f_{i},
\qquad \text{in} \ \Omega,
\end{equation}
\begin{equation}
-\bm{n}\bm{\cdot}\bm{J}_{i} = j_{i}, \qquad \text{on} \ \partial\Omega,
\end{equation}
\end{subequations}
where the flux, $\bm{J}_{i}$, is related to the diffusion and drift, $f_{i}$
is the net rate of production of the ion species per unit volume due to chemical
reaction, and $j_{i}$ is the surface flux on the boundary.

A brush represents a set of macromolecules in a solvent tethered to a surface
that can be characterized as having variation only in the direction perpendicular
to the surface. Therefore, we can make the simplifying assumption that the electric
potential and the concentrations vary only in the $x$-direction to yield
\begin{subequations}\label{poisson_eqn_wbound}
\begin{equation}\label{poisson_eqn}
\frac{\partial}{\partial x}\left(-\varepsilon\frac{\partial\phi}{\partial x}\right)
= \sum_{i} z_{i}N_{A}e\left[C_{i}\right], \qquad \text{for} \ x_{l}<x<x_{r},
\end{equation}
\begin{equation}
-\varepsilon\frac{\partial\phi}{\partial x}\bigg|_{x=x_{l}} = \sigma_{l},
\end{equation}
\begin{equation}
\varepsilon\frac{\partial\phi}{\partial x}\bigg|_{x=x_{r}} = \sigma_{r},
\end{equation}
and
\end{subequations}
\begin{subequations}\label{cons_eq_wbound}
\begin{equation}\label{cons_eq}
\frac{\partial \left[C_{i}\right]}{\partial t} + \frac{\partial J_{i}}{\partial x}
= f_{i}, \qquad \text{for} \ x_{l}<x<x_{r},
\end{equation}
\begin{equation}
J_{i}\bigg|_{x=x_{l}} = j_{i,l},
\end{equation}
\begin{equation}
-J_{i}\bigg|_{x=x_{r}} = j_{i,r},
\end{equation}
\end{subequations}
where $x_{l}$ and $x_{r}$ are the left and right boundaries, respectively, of the domain.

Flux $J_{i}$ depends on the drift velocity, $v_{d,i}$ and the diffusion coefficient, $D_{i}$, by
\begin{equation}
J_{i} = v_{d,i}\left[C_{i}\right] - D_{i}\frac{\partial \left[C_{i}\right]}{\partial x}.
\end{equation}
The drift velocity is related to the external force on the species via the mobility,
which by a modified Stokes-Einstein relation is
\begin{equation}
v_{d,i} = \frac{D_{i}}{\alpha_{i}k_{B}T}F_{i},
\end{equation}
where $k_{B}$ and $T$ are Boltzmann constant and temperature, respectively, and
$\alpha_{i}$ is a positive dimensionless parameter, such that when
$\alpha_{i}=1$, we have the usual Stokes-Einstein relation. Here, the external
force, $F_{i}$, is comprised of the Coulombic force that drives
ion electrophoresis and is proportional to the electric field, and the Born force
that drives mobile ions to regions of higher permittivity. Assuming electrostatic
conditions, this force can be written as
\begin{equation}\label{force_pot}
F_{i} = -z_{i}e\frac{\partial\phi}{\partial x} - \frac{z_{i}^{2}e^{2}}{8\pi r_{i}}
\frac{\partial}{\partial x}\left(\frac{1}{\varepsilon}\right).
\end{equation}
Combining equations (\ref{cons_eq_wbound})--(\ref{force_pot}) yields
\begin{widetext}
\begin{subequations}\label{mod_nernst_planck_wbound}
\begin{equation}\label{mod_nernst_planck}
\frac{\partial \left[C_{i}\right]}{\partial t}
+ \frac{\partial}{\partial x}\left[-D_{i}\frac{\partial\left[C_{i}\right]}{\partial x}
- \frac{D_{i}}{\alpha_{i}k_{B}T}\left(z_{i}e\frac{\partial\phi}{\partial x}
+\frac{z_{i}^{2}e^{2}}{8\pi r_{i}}\frac{\partial}{\partial x}
\left(\frac{1}{\varepsilon}\right)\right)\left[C_{i}\right]\right] = f_{i},
\qquad \text{for} \ x \in \left[x_{l}, x_{r}\right],
\end{equation}
\begin{equation}
\left[-D_{i}\frac{\partial\left[C_{i}\right]}{\partial x}
- \frac{D_{i}}{\alpha_{i}k_{B}T}\left(z_{i}e\frac{\partial\phi}{\partial x}
+\frac{z_{i}^{2}e^{2}}{8\pi r_{i}}\frac{\partial}{\partial x}
\left(\frac{1}{\varepsilon}\right)\right)\left[C_{i}\right]\right]_{x=x_{l}}
= j_{i,l},
\end{equation}
\begin{equation}
\left[D_{i}\frac{\partial\left[C_{i}\right]}{\partial x}
+ \frac{D_{i}}{\alpha_{i}k_{B}T}\left(z_{i}e\frac{\partial\phi}{\partial x}
+\frac{z_{i}^{2}e^{2}}{8\pi r_{i}}\frac{\partial}{\partial x}
\left(\frac{1}{\varepsilon}\right)\right)\left[C_{i}\right]\right]_{x=x_{r}}
= j_{i,r},
\end{equation}
\end{subequations}
\end{widetext}
which is a Born modified form of the Nernst-Planck equation. Together, equations
(\ref{poisson_eqn_wbound}) and (\ref{mod_nernst_planck_wbound}) form the basis
of our MPNP model.

To obtain a dimensionless form, we define the following scaling parameters. We
scale all concentrations by some reference concentration value, $C_{0}$, such that
\begin{equation}
c_{i} \equiv \frac{\left[C_{i}\right]}{C_{0}}.
\end{equation}
Scale the electric potential, $\phi$, with the thermal voltage, $\frac{k_{B}T}{e}
= \frac{N_{A}k_{B}T}{N_{A}e} = \frac{RT}{F}$,
\begin{equation}
y \equiv \frac{\phi}{\left(RT/F\right)},
\end{equation}
where $R$ and $F$ are the gas constant and Faraday's constant, respectively.
Let $\varepsilon = \varepsilon_{0}\varepsilon_{r}\varepsilon_{1}$, where
$\varepsilon_{0}$ is the vacuum permittivity, $\varepsilon_{r}$ is the dielectric
constant of a reference medium and $\varepsilon_{1}$ is the varying portion of the
dimensionless permittivity. We can define the same modified Debye length from
\cite{ceely2023mathematical},
$\lambda_{D}^{2} = \frac{\varepsilon_{0}\varepsilon_{r}RT}{F^{2}C_{0}}$, and
use it to scale the $x$ coordinate and Born radii, $r_{i}$,
\begin{equation}
\hat{x} \equiv \frac{x}{\lambda_{D}}, \quad \hat{r}_{i}\equiv \frac{r_{i}}{\lambda_{D}}.
\end{equation}
Next define a dimensionless Born energy term as
\begin{equation}
\hat{u} \equiv \frac{e^{2}}{8\pi k_{B}T\varepsilon_{0}\varepsilon_{r}\lambda_{D}}.
\end{equation}
Scale all diffusion constants with some value,
$D_{0}$, and define the dimensionless diffusion constants as
\begin{equation}
d_{i} \equiv \frac{D_{i}}{D_{0}}.
\end{equation}
We can then define a time scale using $\lambda_{D}$ and $D_{0}$ to obtain
\begin{equation}
\hat{t} \equiv \frac{t}{(\lambda_{D}^{2}/D_{0})}.
\end{equation}
Finally, the net rate of production is nondimensionalized as
\begin{equation}
\hat{f}_{i} \equiv \frac{\lambda_{D}^{2}}{D_{0}C_{0}}f_{i}.
\end{equation}
This yields the dimensionless equations
\begin{widetext}
\begin{subequations}
\begin{equation}
\frac{\partial}{\partial\hat{x}}\left[-\varepsilon_{1}\frac{\partial y}{\partial\hat{x}}\right]
= \sum_{i} z_{i}c_{i},
\end{equation}
\begin{equation}
\frac{\partial c_{i}}{\partial\hat{t}}
+\frac{\partial}{\partial\hat{x}}\left[-d_{i}\frac{\partial c_{i}}{\partial\hat{x}}
-\frac{d_{i}}{\alpha_{i}}\left(z_{i}\frac{\partial y}{\partial\hat{x}}
+\frac{z_{i}^{2}\hat{u}}{\hat{r}_{i}}\frac{\partial}{\partial\hat{x}}
\left(\frac{1}{\varepsilon_{1}}\right)\right)c_{i}\right] = \hat{f}_{i}.
\end{equation}
\end{subequations}
\end{widetext}
For the boundary conditions, the surface charge density and surface flux are scaled as follows
\begin{equation}
\hat{\sigma} \equiv \frac{\lambda_{D} F}{\varepsilon_{0}\varepsilon_{r}RT}\sigma,
\quad \hat{j} \equiv \frac{\lambda_{D}}{D_{0}C_{0}}j.
\end{equation}

In systems with monovalent ions, the net rates of production are governed by the
binding chemical reactions
\begin{equation}
f_{i} = \sum_{j} -k_{+ij}\left[C_{i}^{+}\right]\left[C_{j}^{-}\right]+k_{-ij}\left[C_{i}C_{j}\right],
\end{equation}
for the production of unbound cation $C_{i}^{+}$,
\begin{equation}
f_{j} = \sum_{i} -k_{+ij}\left[C_{i}^{+}\right]\left[C_{j}^{-}\right]+k_{-ij}\left[C_{i}C_{j}\right],
\end{equation}
for the production of unbound anion $C_{j}^{-}$, and
\begin{equation}
f_{ij} = k_{+ij}\left[C_{i}^{+}\right]\left[C_{j}^{-}\right]-k_{-ij}\left[C_{i}C_{j}\right],
\end{equation}
for the production of bound cation-anion pair $C_{i}C_{j}$.

The forward reaction rate constants, $k_{+ij}$, can be nondimensionalized as
\begin{equation}
\hat{k}_{+ij} \equiv \frac{\lambda_{D}^{2}C_{0}}{D_{0}}k_{+ij}.
\end{equation}
The backward reaction rate constants, $k_{-ij}$, can be nondimensionalized as
\begin{equation}
\hat{k}_{-ij} \equiv \frac{\lambda_{D}^{2}}{D_{0}}k_{-ij}.
\end{equation}

Considering a negatively charged GAG brush, $g$, (from $\hat{x}=0$ to
$\hat{x}=\hat{\ell}=\ell/\lambda_{D}$) in a monovalent salt solution (from
$\hat{x}=\hat{\ell}$ to $\hat{x}=\hat{L}=L/\lambda_{D}$)
consisting of one cation, $c$, and one anion, $a$, the MPNP model equations can
be written as 
\begin{widetext}
\begin{subequations}\label{mpnp_model}
\begin{equation}\label{pde_pot}
\frac{\partial}{\partial\hat{x}}\left[-\varepsilon_{1}\frac{\partial y}{\partial\hat{x}}\right]
= c - a - g,
\end{equation}
\begin{equation}\label{pde_cat}
\frac{\partial c}{\partial\hat{t}}
+\frac{\partial}{\partial\hat{x}}\left[-d_{c}\frac{\partial c}{\partial\hat{x}}
-\frac{d_{c}}{\alpha_{c}}\left(\frac{\partial y}{\partial\hat{x}}
+\frac{\hat{u}}{\hat{r}_{c}}\frac{\partial}{\partial\hat{x}}
\left(\frac{1}{\varepsilon_{1}}\right)\right)c\right]
=-\hat{k}_{+}cg+\hat{k}_{-}\left[cg\right],
\end{equation}
\begin{equation}\label{pde_an}
\frac{\partial a}{\partial\hat{t}}
+\frac{\partial}{\partial\hat{x}}\left[-d_{a}\frac{\partial a}{\partial\hat{x}}
-\frac{d_{a}}{\alpha_{a}}\left(-\frac{\partial{y}}{\partial\hat{x}}
+\frac{\hat{u}}{\hat{r}_{a}}\frac{\partial}{\partial\hat{x}}
\left(\frac{1}{\varepsilon_{1}}\right)\right)a\right]
=0,
\end{equation}
\begin{equation}\label{pde_brush}
\frac{\partial g}{\partial\hat{t}} = -\hat{k}_{+}cg+\hat{k}_{-}\left[cg\right],
\end{equation}
\begin{equation}\label{pde_bound}
\frac{\partial \left[cg\right]}{\partial\hat{t}}
= \hat{k}_{+}cg-\hat{k}_{-}\left[cg\right].
\end{equation}
\end{subequations}
\end{widetext}
Equations (\ref{pde_brush}) and (\ref{pde_bound}) represent the time evolution
of the concentrations of the
unbound GAG brush and the bound cation-brush pair. The GAG brush is fixed, and
therefore, it is not mobile. Thus the diffusivity and drift terms, and
consequently the flux term, disappear. For this research, permittivity is assumed
to vary spatially as $\varepsilon_{1}(\hat{x}) = \frac{1}{2}(\varepsilon_{G}
-\varepsilon_{S})\left[\tanh\left(\frac{1-{\hat{x}}/{\hat{\ell}}}{\beta}\right)
+1\right] + \varepsilon_{S}$, where $\varepsilon_{G}$ and $\varepsilon_{S}$ are
the values of the dimensionless permittivity in the brush (at $\hat{x}=0$) and
in the salt (at $\hat{x}=\hat{L}$), respectively, and $\beta$ is a smoothing
parameter to control the slope of the transition.

For our systems of interest, there are no surface charge densities
on either boundary and no sources or sinks for the cation or anion, thus the
boundary conditions for all of the independent variables reduce to homogeneous
Neumann boundary conditions:
\begin{equation}\label{mpnp_boundary}
\frac{\partial u}{\partial \hat{x}}(0) = 0, \quad
\frac{\partial u}{\partial \hat{x}}(\hat{L}) = 0,
\end{equation}
where $u$ represents any of the independent variables $y$, $c$, $a$, $g$, or
$\left[cg\right]$.

\subsection{Energy functional}\label{energy-functional}
The system can alternatively be described by an energy functional, and the model
equations can be derived via a variational approach. Using the mean-field
approximation, the energy functional is of the form
\begin{equation}\label{orig_energy_func}
\mathcal{F} = \int_{\Omega}\bigg\{-\frac{\varepsilon_{0}\varepsilon_{r}\varepsilon_{1}}{2}|\nabla\phi|^{2}
+ \sum_{i}\left[C_{i}\right](\mu_{i} - N_{A}k_{B}T)\bigg\} \ dV,
\end{equation}
where $\mu_{i}$ is the chemical potential for ionic species, $i$. The chemical
potential consists of electric potential, entropic, and Born solvation energy
\cite{liu2017incorporating} terms such that
\begin{eqnarray}
\mu_{i} = \mu_{i}^{0} &+& z_{i}eN_{A}\phi
+ N_{A}k_{B}T\ln{\left(\frac{\left[C_{i}\right]}{C_{0}}\right)}\nonumber\\
&+& N_{A}\frac{z_{i}^{2}e^{2}}{8\pi \varepsilon_{0}\varepsilon_{r}r_{i}}
\left(\frac{1}{\varepsilon_{1}} - \frac{1}{\varepsilon_{S}}\right),
\end{eqnarray}
where $\varepsilon_{S}$ is the value of $\varepsilon_{1}$ deep in the salt side.
Focusing on the system represented by equations (\ref{mpnp_model}), we have
\begin{subequations}\label{orig_chem_pot}
\begin{eqnarray}
\mu_{c} = \mu_{c}^{0} &+& eN_{A}\phi
+ N_{A}k_{B}T\ln{\left(\frac{\left[C^{+}\right]}{C_{0}}\right)}\nonumber\\
&+& N_{A}\frac{e^{2}}{8\pi \varepsilon_{0}\varepsilon_{r}r_{c}}
\left(\frac{1}{\varepsilon_{1}} - \frac{1}{\varepsilon_{S}}\right),
\end{eqnarray}
\begin{eqnarray}
\mu_{a} = \mu_{a}^{0} &-& eN_{A}\phi
+ N_{A}k_{B}T\ln{\left(\frac{\left[A^{-}\right]}{C_{0}}\right)}\nonumber\\
&+& N_{A}\frac{e^{2}}{8\pi \varepsilon_{0}\varepsilon_{r}r_{a}}
\left(\frac{1}{\varepsilon_{1}} - \frac{1}{\varepsilon_{S}}\right),
\end{eqnarray}
\begin{equation}
\mu_{g} = \mu_{g}^{0} - eN_{A}\phi
+ N_{A}k_{B}T\ln{\left(\frac{\left[G^{-}\right]}{C_{0}}\right)},
\end{equation}
\begin{equation}
\mu_{cg} = \mu_{cg}^{0}
+ N_{A}k_{B}T\ln{\left(\frac{\left[CG\right]}{C_{0}}\right)}.
\end{equation}
\end{subequations}
The total chemical potential energy is
\begin{equation}
E = \left[C^{+}\right]\mu_{c} + \left[A^{-}\right]\mu_{a} + \left[G^{-}\right]\mu_{g}
+ \left[CG\right]\mu_{cg}.
\end{equation}
If we define $\eta$ as the extent of the reaction $C^{+} + G^{-} \to CG$, and
substitute $\left[C^{+}\right] = C^{0} - \eta$, $\left[G^{-}\right] = G^{0} - \eta$ and
$\left[CG\right] = \eta$ into the total chemical potential energy, we can minimize
$E$ to find the equilibrium state by finding where $\frac{\partial E}{\partial \eta} = 0$.
\begin{widetext}
\begin{equation}
\begin{aligned}
\frac{\partial E}{\partial \eta} &= -\left(\mu_{c}^{0} + \mu_{g}^{0} - \mu_{cg}^{0}\right)
+ N_{A}k_{B}T\left[\ln{\left(\frac{\left[CG\right]}{C_{0}}\right)}
- \ln{\left(\frac{\left[C^{+}\right]}{C_{0}}\right)}
- \ln{\left(\frac{\left[G^{-}\right]}{C_{0}}\right)}\right]
- N_{A}k_{B}T
- N_{A}\frac{e^{2}}{8\pi \varepsilon_{0}\varepsilon_{r}r_{c}}
\left(\frac{1}{\varepsilon_{1}} - \frac{1}{\varepsilon_{S}}\right)\\
&= -\left(\mu_{c}^{0} + \mu_{g}^{0} - \mu_{cg}^{0}\right)
+ N_{A}k_{B}T\ln{\left(\frac{\left[CG\right]C_{0}}{\left[C^{+}\right]\left[G^{-}\right]}\right)}
- N_{A}k_{B}T
- N_{A}\frac{e^{2}}{8\pi \varepsilon_{0}\varepsilon_{r}r_{c}}
\left(\frac{1}{\varepsilon_{1}} - \frac{1}{\varepsilon_{S}}\right) = 0.
\end{aligned}
\end{equation}
\end{widetext}
Considering $x = 0$, we have $\varepsilon_{1} = \varepsilon_{G}$. At equilibrium,
the dissociation constant, which is the ratio of the backward
and forward reaction rate constants and has units of concentration, is equal to
$K = \frac{\left[C^{+}\right]\left[G^{-}\right]}{\left[CG\right]}$. This yields
\begin{eqnarray}
\mu_{c}^{0} + \mu_{g}^{0} - \mu_{cg}^{0}
= & &N_{A}k_{B}T\left[\ln{\left(\frac{C_{0}}{K}\right)} - 1\right]\nonumber\\
&-& N_{A}\frac{e^{2}}{8\pi \varepsilon_{0}\varepsilon_{r}r_{c}}
\left(\frac{1}{\varepsilon_{G}} - \frac{1}{\varepsilon_{S}}\right).
\end{eqnarray}
Choosing reference values $\mu_{c}^{0} = \mu_{a}^{0} = \mu_{g}^{0} = 0$, we have
\begin{equation}
\mu_{cg}^{0}
= N_{A}k_{B}T\left[1 + \ln{\left(\frac{K}{C_{0}}\right)} + \hat{U}\right],
\end{equation}
where $\hat{U} = \frac{e^{2}}{8\pi k_{B}T\varepsilon_{0}\varepsilon_{r}r_{c}}
\left(\frac{1}{\varepsilon_{G}} - \frac{1}{\varepsilon_{S}}\right)$.
Putting all of this together, equations (\ref{orig_chem_pot}) become
\begin{subequations}\label{chem_pot}
\begin{eqnarray}
\mu_{c} = eN_{A}\phi
&+& N_{A}k_{B}T\ln{\left(\frac{\left[C^{+}\right]}{C_{0}}\right)}\nonumber\\
&+& N_{A}\frac{e^{2}}{8\pi \varepsilon_{0}\varepsilon_{r}r_{c}}
\left(\frac{1}{\varepsilon_{1}} - \frac{1}{\varepsilon_{S}}\right),
\end{eqnarray}
\begin{eqnarray}
\mu_{a} = -eN_{A}\phi
&+& N_{A}k_{B}T\ln{\left(\frac{\left[A^{-}\right]}{C_{0}}\right)}\nonumber\\
&+& N_{A}\frac{e^{2}}{8\pi \varepsilon_{0}\varepsilon_{r}r_{a}}
\left(\frac{1}{\varepsilon_{1}} - \frac{1}{\varepsilon_{S}}\right),
\end{eqnarray}
\begin{equation}
\mu_{g} = -eN_{A}\phi
+ N_{A}k_{B}T\ln{\left(\frac{\left[G^{-}\right]}{C_{0}}\right)},
\end{equation}
\begin{equation}
\mu_{cg} = N_{A}k_{B}T\left[\ln{\left(\frac{\left[CG\right]}{C_{0}}\right)}
+ \ln{\left(\frac{K}{C_{0}}\right)} + 1 + \hat{U}\right].
\end{equation}
\end{subequations}


As before, we make the simplifying assumption that there is only variation in the
$x$-direction. Thus, we can write an energy density functional as
$\mathcal{F}/\mathcal{A}$
where $\mathcal{A}$ is the cross-sectional area in the $y$ and $z$ dimensions.
We can scale the energy density functional by $N_{A}k_{B}TC_{0}\lambda_{D}$ to define
a dimensionless energy density functional as
\begin{equation}\label{energy_scale}
\hat{\mathcal{F}} = \frac{\mathcal{F}/\mathcal{A}}{N_{A}k_{B}TC_{0}\lambda_{D}}.
\end{equation}

Substituting equations (\ref{chem_pot}) and (\ref{energy_scale}) into equation (\ref{orig_energy_func}) and using all previously
defined scaling parameters, we can express the dimensionless energy density
functional as
\begin{widetext}
\begin{equation}\label{nondim_energy_dens_func}
\begin{aligned}
\hat{\mathcal{F}} = \int_{\Omega_{x}}\bigg\{&-\frac{\varepsilon_{1}}{2}
\left(\frac{\partial y}{\partial\hat{x}}\right)^{2} + \left(c-a-g\right)y
+ \bigg[c\left(\ln{c}-1\right) + a\left(\ln{a}-1\right) + g\left(\ln{g}-1\right)
+ [cg]\left(\ln{[cg]}+\ln{\widetilde{K}} + \hat{U}\right)\bigg]\\
&+ \left[\frac{c}{\hat{r}_{c}} + \frac{a}{\hat{r}_{a}}\right]\hat{u}
\left(\frac{1}{\varepsilon_{1}} - \frac{1}{\varepsilon_{S}}\right)\bigg\} \ d\hat{x}.
\end{aligned}
\end{equation}
\end{widetext}
This can be written as the sum of four separate energy density terms, which are the
electrostatic field distribution energy density
\[\left(\hat{\mathcal{F}}_{1}=\int-\frac{\varepsilon_{1}}{2}
\left(\frac{\partial y}{\partial\hat{x}}\right)^{2} \ d\hat{x} \right),\]
the electrostatic potential energy density
\[\left(\hat{\mathcal{F}}_{2}=\int\left(c-a-g\right)y \ d\hat{x}\right),\]
the entropic/binding energy density
\begin{eqnarray*}
\bigg(\hat{\mathcal{F}}_{3}=\int\bigg[c\left(\ln{c}-1\right) &+& a\left(\ln{a}-1\right)
+ g\left(\ln{g}-1\right)\\
&+& [cg]\left(\ln{[cg]}+\ln{\widetilde{K}}+\hat{U}\right)\bigg]
\ d\hat{x}\bigg),
\end{eqnarray*}
and the Born solvation energy density
\[\left(\hat{\mathcal{F}}_{4}=\int\left[\frac{c}{\hat{r}_{c}} + \frac{a}{\hat{r}_{a}}\right]\hat{u}
\left(\frac{1}{\varepsilon_{1}} - \frac{1}{\varepsilon_{S}}\right) \ d\hat{x}\right).\]

\section{Results and discussions}\label{results}
\subsection{Initially far from equilibrium}\label{init_far}
To test whether the proposed model (\ref{mpnp_model})-(\ref{mpnp_boundary}) approaches
the same steady state solutions of the MPB model, we compared numerical results
obtained using finite differences
to the results of the four GAG Brush/salt systems from
\cite{ceely2023mathematical} and \cite{sterling2021ion}, Hyaluronic Acid (HA)-NaCl,
HA-KCl, Heparan Sulfate (HS)-NaCl, and HS-KCl. All necessary parameters
that impact the steady state solutions were derived from \cite{ceely2023mathematical}.
The parameters $d_{c}$, $d_{a}$, $\hat{k}_{+}$, and $\hat{k}_{-}$ as well as the
initial conditions impact the transient response, but do not affect the steady
state solution and could not be directly derived from either \cite{ceely2023mathematical} or
\cite{sterling2021ion}. The molecular dynamics simulations in \cite{sterling2021ion}
reached quasi-steady state in about 100 ns. We chose a time scale of approximately
0.5 ns such that this corresponds to approximately 200
dimensionless time units. Thus, we chose $D_{0} = 1.5\times 10^{-9} \ m^{2}/s$.
To make $d_{c}$ and $d_{a}$ slightly more and slightly less than 1, respectively,
such that the diffusion occurred on this time scale,
we chose $D_{c} = 1.75\times 10^{-9}$ and $D_{a} = 1.25\times 10^{-9} \ m^{2}/s$.
The dissociation constant, $\widetilde{K}$, and thus the ratio 
$\hat{k}_{-}/\hat{k}_{+}$ are known. We chose reasonable values for $\hat{k}_{+}$
and $\hat{k}_{-}$ such that the binding occurred on the same time scale or faster. The
complete list of parameters and the values used are contained in Table~\ref{input_params}.
The supporting information for \cite{sterling2021ion} provides the charge of the
GAG brushes and the number of each ion used for the four different scenarios.

\begin{table}
  \caption{Input Parameters to MPNP Model.}
  \label{input_params}
  \begin{ruledtabular}
  \begin{tabular}{lllll}
    \hline
    \textbf{Parameter} & \textbf{HA-}
    & \textbf{HA-} & \textbf{HS-} & \textbf{HS-} \\
     & \textbf{NaCl} & \textbf{KCl}
     & \textbf{NaCl} & \textbf{KCl} \\
    \hline
    $\hat{\ell}$ & $7.838$ & $8.206$ & $7.079$ & $6.592$ \\
    $\hat{L}$ & $29.007$ & $29.007$ & $28.484$ & $27.951$ \\
    $\varepsilon_{S}$ & $0.773$ & $0.791$ & $0.756$ & $0.774$ \\
    $\varepsilon_{G}$ & $0.649$ & $0.655$ & $0.480$ & $0.485$ \\
    $\beta$ & 0.1 & 0.1 & 0.1 & 0.1 \\
    $\hat{u}$ & 0.417 & 0.417 & 0.409 & 0.401 \\
    $\hat{r}_{c}$ & 0.196 & 0.236 & 0.192 & 0.227 \\
    $d_{c}$ & 1.167 & 1.167 & 1.167 & 1.167 \\
    $\alpha_{c}$ & 1 & 1 & 1 & 1 \\
    $\hat{r}_{a}$ & 0.0273 & 0.0273 & 0.0268 & 0.0263 \\
    $d_{a}$ & 0.833 & 0.833 & 0.833 & 0.833 \\
    $\alpha_{a}$ & 1 & 1 & 1 & 1 \\
    $\hat{k}_{+}$ & 1 & 1 & 9.351 & 143.097 \\
    $\hat{k}_{-}$ & 0.172 & 0.114 & 0.125 & 0.125 \\
    \hline
  \end{tabular}
  \end{ruledtabular}
\end{table}

\begin{figure*}
  \includegraphics[scale=0.40]{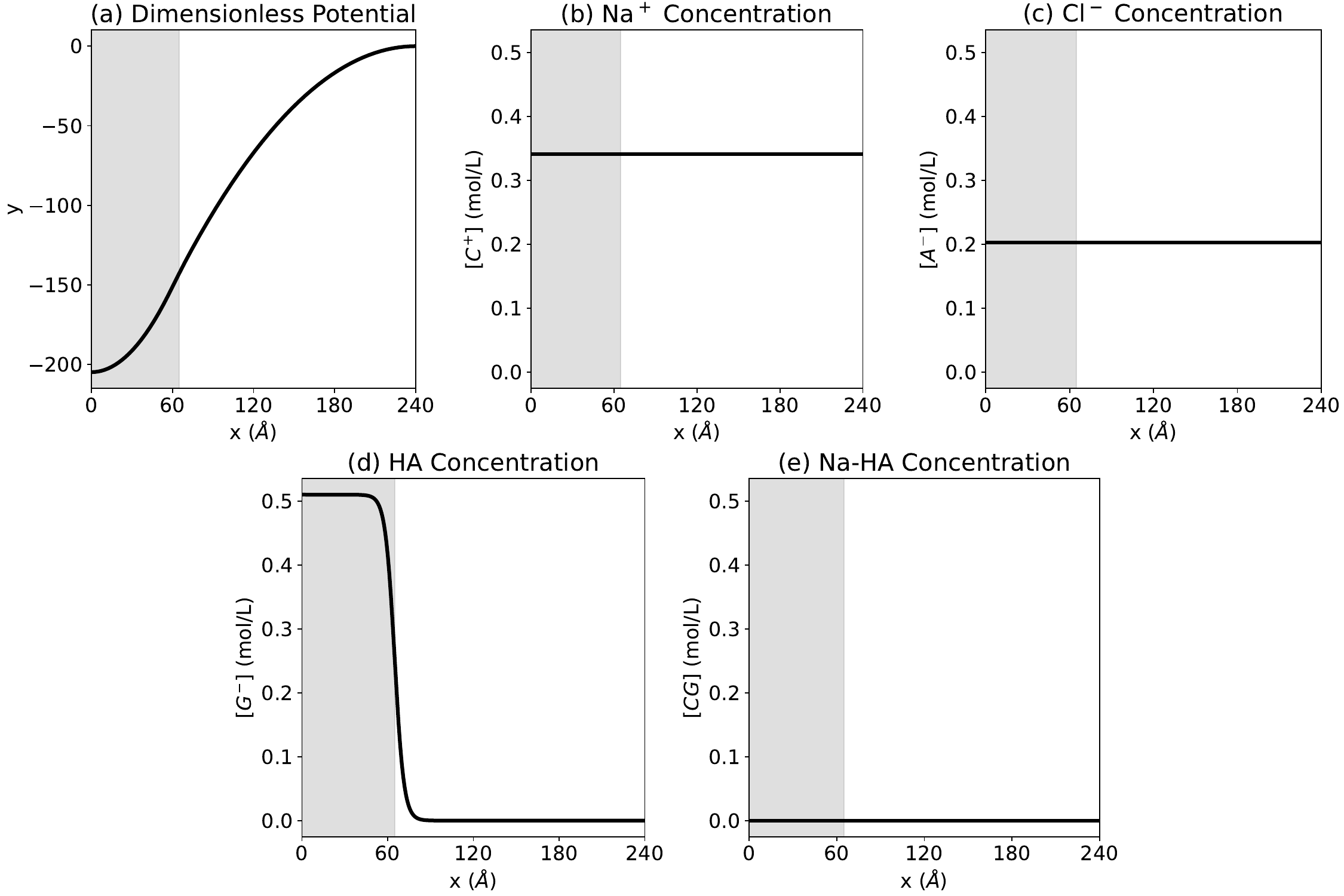}
  \caption{Initial conditions for the HA-NaCl system
  far from equilibrium. The shaded region represents the brush region.}
  \label{HANaCl_init_cond}
\end{figure*}

For the initial conditions, 
we assumed the ions to be uniformly distributed and that the GAG brush and cation
are initially completely unbound. These start the systems sufficiently far from
equilibrium, but do not represent physically realistic initial conditions. However,
starting far from equilibrium allows us the opportunity to examine in more detail
how diffusion, electrophoresis, the Born force, and ion pairing each play a role
in driving the system towards steady state. See Figure~\ref{HANaCl_init_cond} for
plots of the initial conditions for the HA-NaCl system. The initial conditions
for the other systems have similar characteristics and their plots are not included.
It should be noted that the initial
electric potential is not specified, but rather computed by applying equations
(\ref{pde_pot}) and (\ref{mpnp_boundary}) using the initial conditions for the
concentrations of the unbound cation, anion, and GAG brush.

\begin{figure*}
  \includegraphics[scale=0.40]{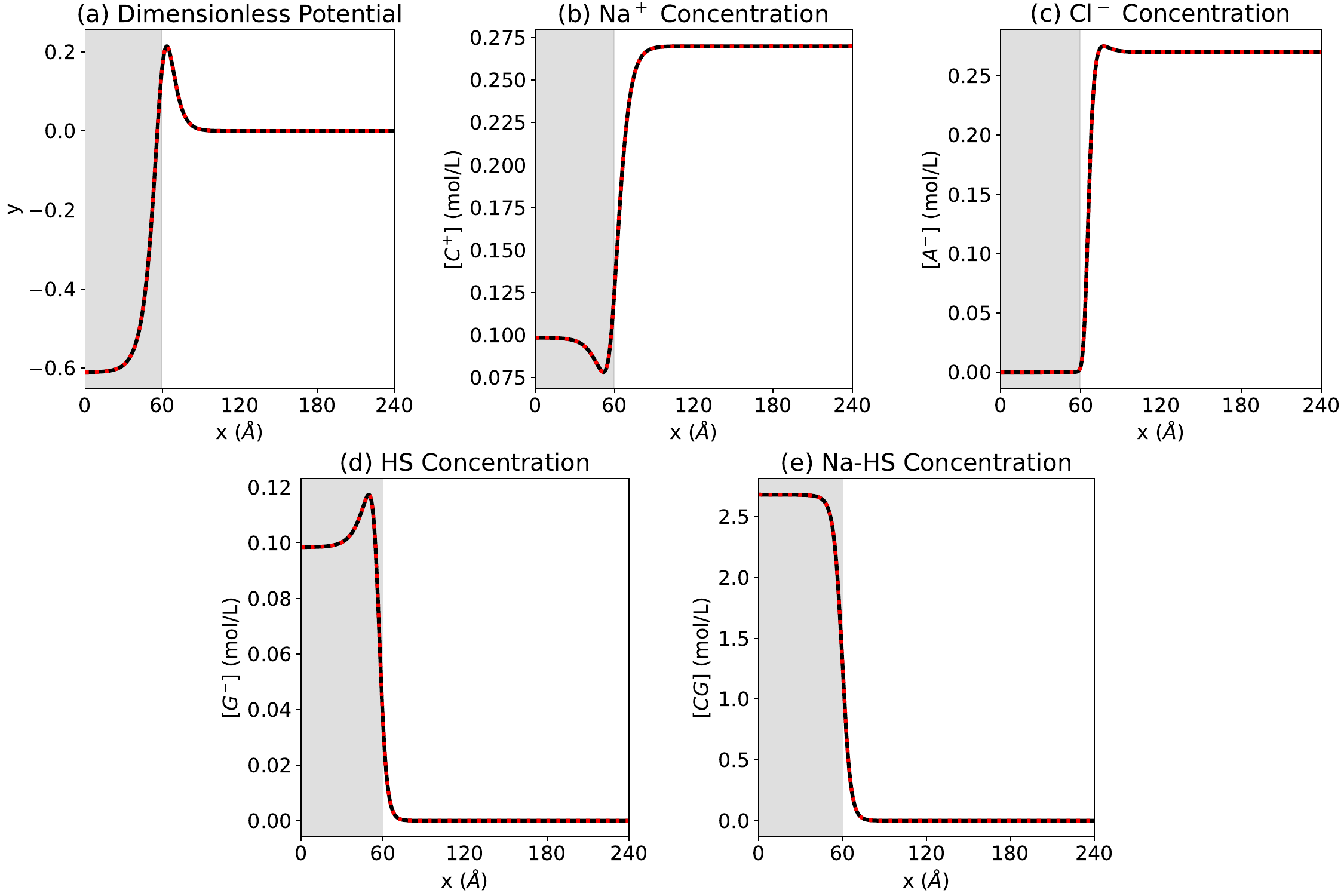}
  \caption{Overlay of the MPB model solution (solid black curve) and the MPNP model
  steady state solution (dotted red curve) for the HS-NaCl system.
  The shaded region represents the brush region.}
  \label{HSNaCl_ss_comp}
\end{figure*}

Using the parameters in Table~\ref{input_params} and the initial conditions described above, 
we ran our proposed model for 400 time units to ensure steady state was achieved. 
We obtained excellent agreement between the MPB model solution and our proposed 
MPNP model steady state solution. Figure~\ref{HSNaCl_ss_comp} contains an overlay
of the results for the HS-NaCl system. As seen in the figure,
the steady state results of the two models are indistinguishable. Overlays of the results
for the other three systems studied are not included, but also show indistinguishable
results between the two models.

The benefit of our proposed MPNP model over the MPB model is the ability to model
the transient behavior of GAG brush/salt systems in addition to the steady state.
While the transient results presented here may not match the true transient
behavior of the four systems according to the molecular dynamics simulations,
the results accurately simulate the model equations for the chosen values of
diffusion coefficients, reaction rate constants, and initial conditions.
The transient behavior of the unbound cation and anion concentrations are driven
by diffusion, electrophoresis, the Born force, and ion pairing (cation only),
whereas the unbound GAG brush and bound cation-brush concentrations are only driven
by ion pairing. When a gradient in the ion concentration exists, the ions will
tend to move from regions of higher to lower concentration due to diffusion.
When a gradient in the electric potential exists, cations will tend to move from
regions of higher to lower electric potential, whereas anions will tend to move
from regions of lower to higher electric potential due to electrophoresis. When
a gradient in permittivity exists, the ions will tend to move from regions of
lower to higher permittivity due to the Born force.

In our application of the MPNP model to the molecular dynamics simulation results
of GAG brushes, 
the permittivity was found to be lower in the brush region, higher in the salt
region and the change in permittivity occurs near the brush/salt interface. Thus,
the Born force acts at the interface and wants to move/keep ions out of the brush
region and move/keep ions into the salt region. Ion pairing is not a movement
of ions, but rather the binding and unbinding of the cation to the GAG brush. When
binding is occurring, there is a decrease in unbound cation and GAG brush
concentrations and an increase in bound cation-brush concentration. Conversely,
when unbinding is occurring, there is an increase in unbound cation and GAG brush
concentrations and a decrease in bound cation-brush concentration. Binding and
unbinding occur simultaneously, however the system can be in a phase of net
binding, unbinding or ion pairing equilibrium depending on the forward and backward
reaction rates and the concentrations. We generated plots of each of these terms
separately to aid in the discussion that follows, but the plots are not included.

\begin{figure*}  
  \includegraphics[scale=0.40]{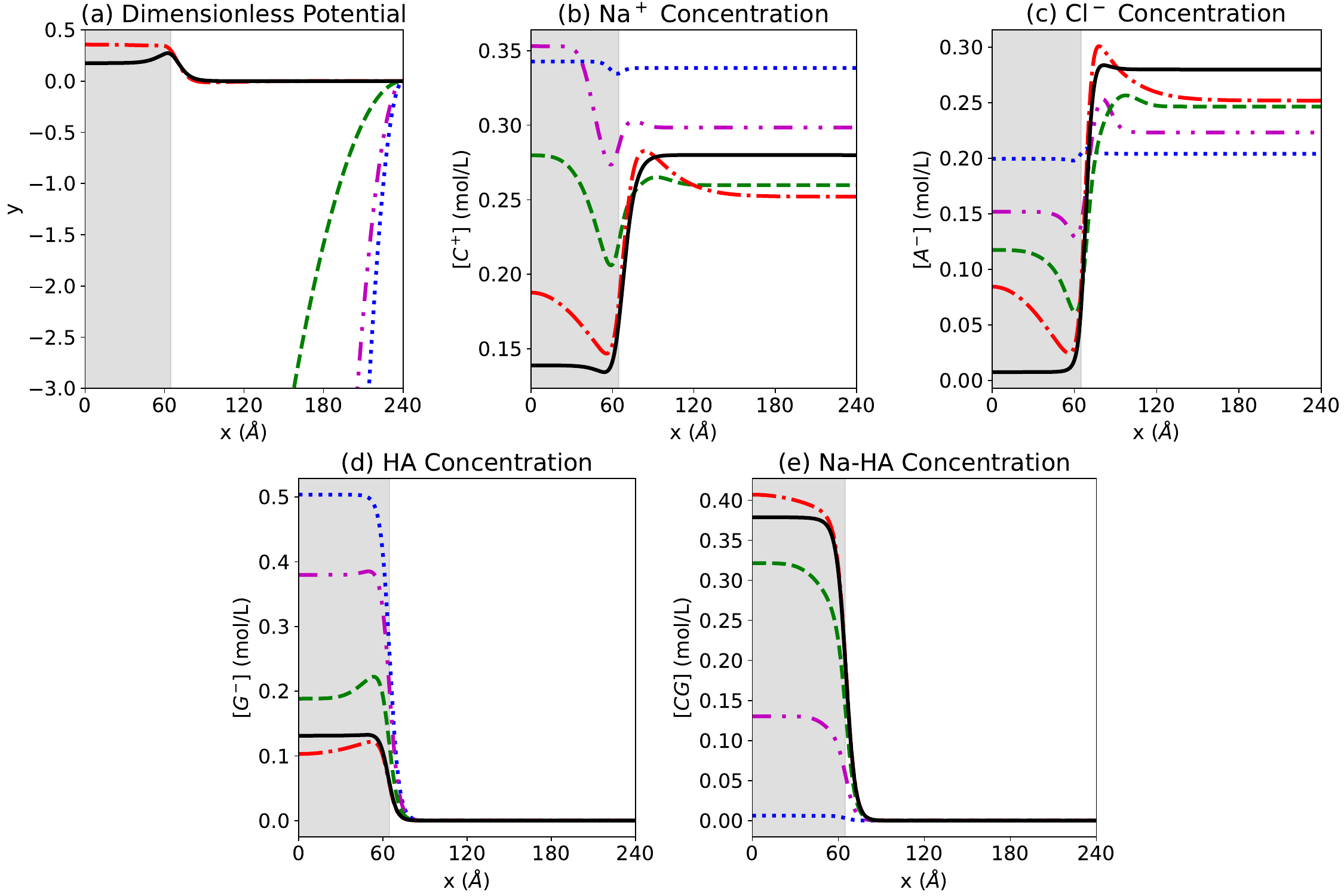}
  \caption{Transient results at $\hat{t}=0.01$ (dotted blue curve), $\hat{t}=0.24$
  (dashed dot dot magenta curve), $\hat{t}=0.97$ (dashed green curve), $\hat{t}=13$
  (dashed dot red curve), and $\hat{t}=384$ (solid black curve) for the HA-NaCl
  system initially far from equilibrium. The shaded region represents
  the brush region.}
  \label{HANaCl_trans}
\end{figure*}

\begin{figure*}
  \includegraphics[scale=0.40]{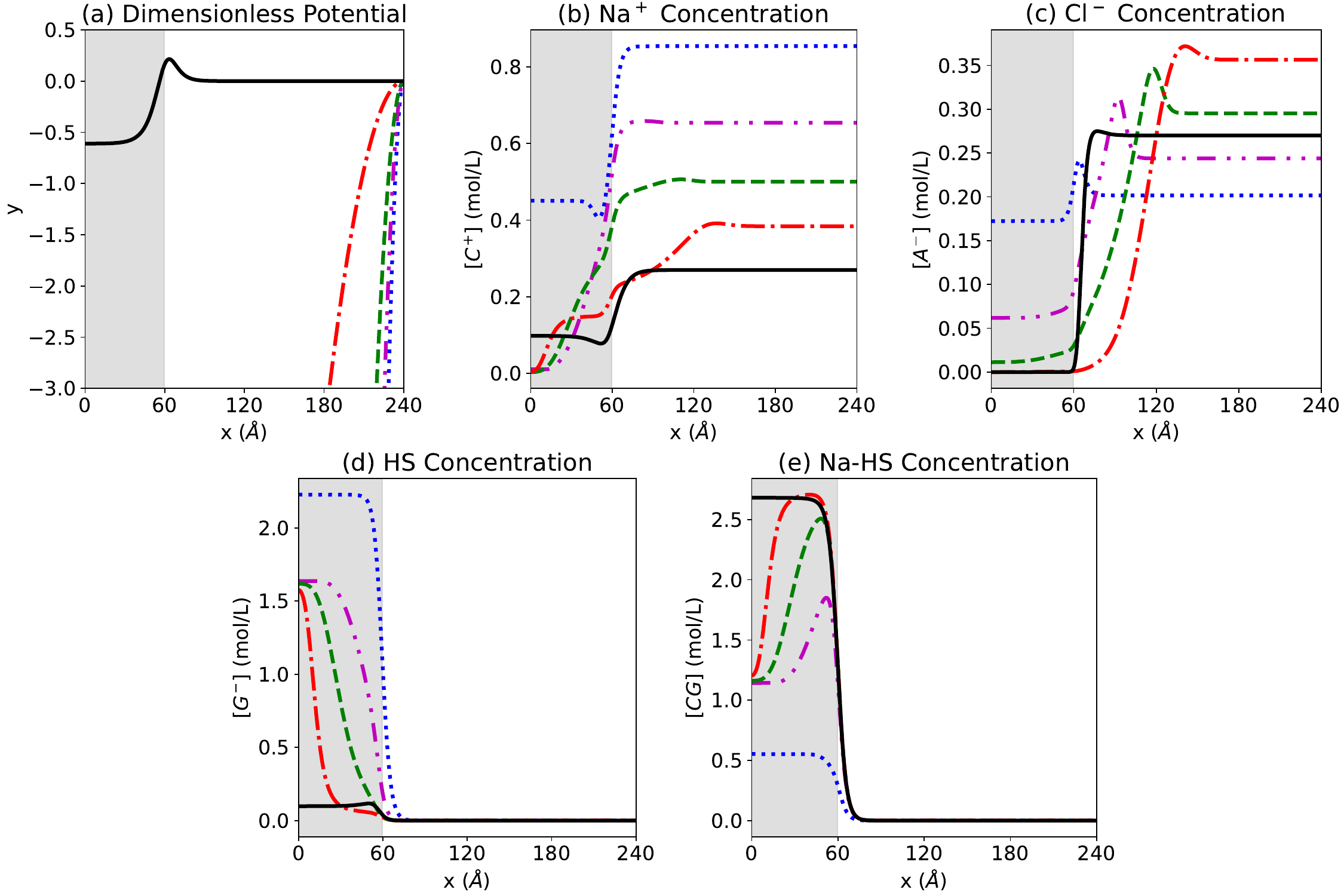}
  \caption{Transient results at $\hat{t}=0.01$ (dotted blue curve), $\hat{t}=0.1$
  (dashed dot dot magenta curve), $\hat{t}=0.26$ (dashed green curve), $\hat{t}=0.79$
  (dashed dot red curve), and $\hat{t}=383$ (solid black curve) for the HS-NaCl
  system initially far from equilibrium. The shaded region represents
  the brush region.}
  \label{HSNaCl_trans}
\end{figure*}

\begin{figure*}
  \includegraphics[scale=0.40]{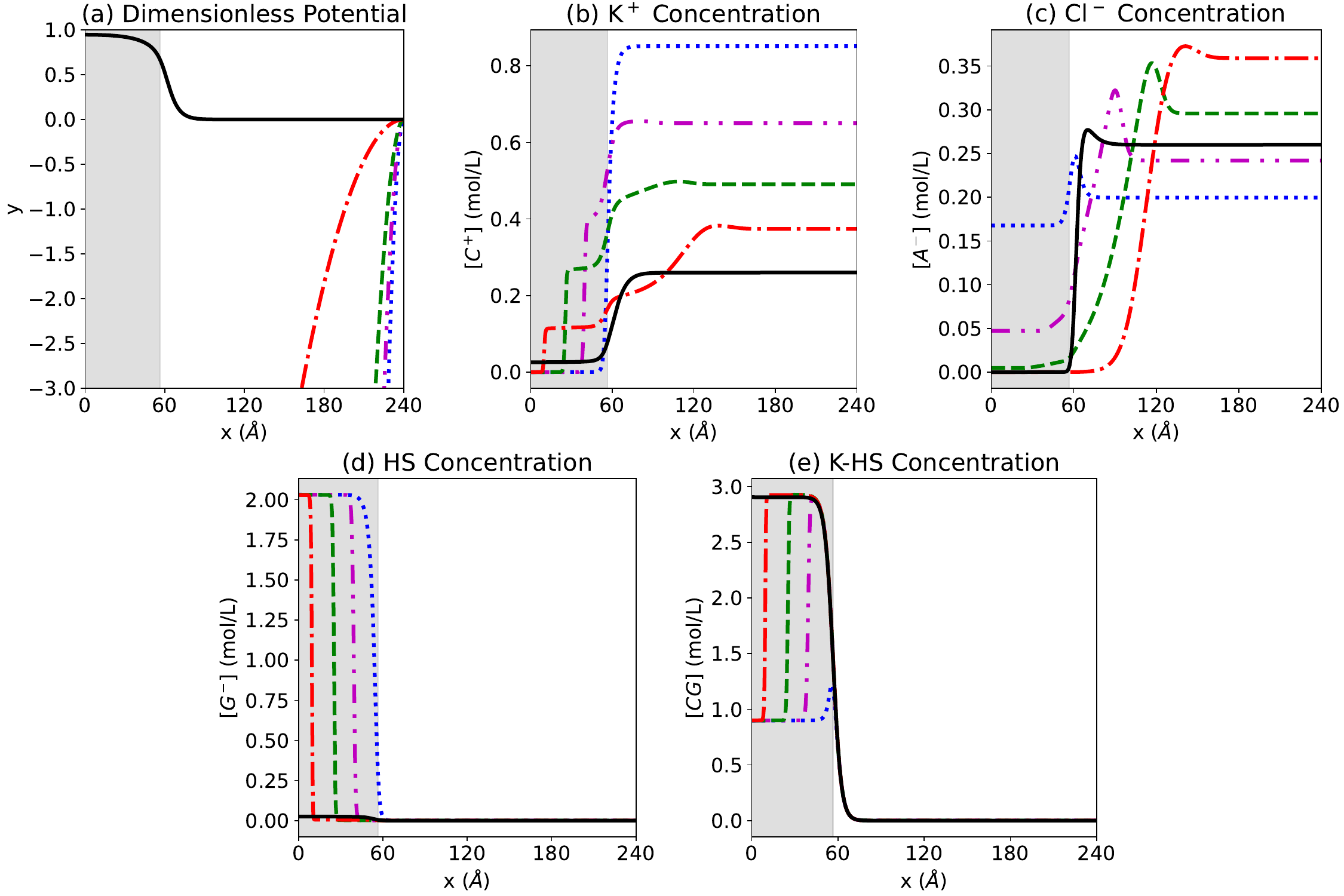}
  \caption{Transient results at $\hat{t}=0.01$ (dotted blue curve), $\hat{t}=0.1$
  (dashed dot dot magenta curve), $\hat{t}=0.27$ (dashed green curve), $\hat{t}=0.95$
  (dashed dot red curve), and $\hat{t}=384$ (solid black curve) for the HS-KCl
  system initially far from equilibrium. The shaded region represents
  the brush region.}
  \label{HSKCl_trans}
\end{figure*}

Figures~\ref{HANaCl_trans}--\ref{HSKCl_trans}
show transient results for the systems at selected time steps.
For all four systems, at the initial time, there is no diffusion occurring.
Electrophoresis is driving cations into the brush region and anions into the salt
region. The Born force is driving both cations and anions at the interface into
the salt region. Only binding can occur initially as both the cation and brush
are completely unbound. The initial strength of each of these driving phenomena
is different for the four systems and is discussed below.

Beginning first with the HA-NaCl and HA-KCl systems, we see that both exhibit similar
transient behaviors. This is the expected behavior as the parameters and
initial conditions for the two systems are similar. Only the results of the
HA-NaCl system plotted in Figure~\ref{HANaCl_trans} are included, however, the
discussion is applicable to both systems.
The net change for the cation concentration is an initial increase in
the brush region (electrophoresis dominates), decrease at the interface
(electrophoresis, Born force, and binding work together), and decrease in the
salt region (electrophoresis dominates).
The net change for the anion concentration is an initial decrease in the brush
region (electrophoresis dominates), decrease to the left of and increase to the
right of the interface (Born dominates), and increase in the salt region
(electrophoresis dominates). The brush concentration decreases and the
bound cation-brush concentration increases slightly.

As time progresses, electrophoresis continues to drive cations into the brush
region and anions into the salt region. However, the electric potential is
increasing with time and therefore decreasing the impact of electrophoresis. 
Diffusion begins to occur around the interface and competes against the Born force.
Net binding continues in the brush region decreasing the unbound cation and brush
concentrations. Around $\hat{t}=2$, the ion pairing overshoots the steady state,
but continues net binding. Around $\hat{t}=3$, the electric potential becomes positive in
the brush region and electrophoresis begins to drive cations into the salt region,
especially around the interface, eventually acting as a barrier in chorus with
the Born force. For the anion, electrophoresis and diffusion work together to
drive anions into the brush region, but are dominated by the Born force, ultimately
driving most of the anions into the salt region. Around $\hat{t}=4$, the electric
potential overshoots the steady state. Around $\hat{t}=5$, the ion pairing
switches to net unbinding and increases the number of mobile unbound cations,
which are driven into the salt region by electrophoresis and the Born force.
Around $\hat{t}=11.5$, the electric potential reaches its maximum in the brush
and begins to decrease towards steady state. Steady state is achieved around
$\hat{t}=384$, at which time ion pairing has reached equilibrium and any cation
or anion movement due to diffusion, electrophoresis or the Born force cancel out
for zero net movement of ions.

The HS-NaCl and HS-KCl systems in Figures~\ref{HSNaCl_trans} and
\ref{HSKCl_trans}, respectively, also share similar parameters and initial
conditions with the exception of the forward reaction rate being approximately
15 times larger for the binding of HS to K compared to Na. Due to this difference,
these systems will be discussed separately.

Focusing on the HS-NaCl system,
the net change for the cation concentration is an initial rapid decrease in
the brush region (binding dominates), decrease at the interface (Born force
and binding dominate), and decrease in the salt region (electrophoresis dominates).
The net change for the anion concentration is an initial decrease in the brush
region (electrophoresis dominates), decrease to the left of and increase to the
right of the interface (Born dominates), and increase in the salt region
(electrophoresis dominates). The brush concentration decreases and
the bound cation-brush concentration increases significantly.

As time progresses, all of the cations initially in the brush become bound effectively
halting any binding at the brush boundary temporarily. Binding continues to occur
at the interface as electrophoresis and diffusion (to a much lesser extent)
drive cations into the brush region. Anions continue their movement from the
brush region and into the salt region as a result of electrophoresis and the Born
force. Once in the salt region, diffusion drives the anions to the steady state
concentration. As the ion pairing reaches equilibrium to the left of the interface,
electrophoresis and diffusion can now move cations further into the brush region
allowing binding to resume at the boundary. This process slows until steady state
is achieved around $\hat{t}=383$, at which time ion pairing has reached equilibrium
throughout the whole brush region and any cation or anion movement due to diffusion,
electrophoresis or the Born force cancel out for zero net movement of ions.

Moving onto the HS-KCl system,
the anion behaves in the exact same manner as previously described for the
HS-NaCl system and no further discussion is provided.
The net change for the cation concentration is an initial near total decrease in
the brush region (binding dominates as all unbound cations are almost instantaneously
bound to the brush) and decrease in the salt region (electrophoresis dominates).
The effects of the Born force are completely masked by binding and electrophoresis.
The brush concentration decreases and the bound cation-brush concentration
increases significantly.

Because all of the cations initially in the brush region were almost instantaneously
bound to the brush, all binding is effectively halted throughout the brush region
except at the interface where cations are driven into the brush region by
electrophoresis and diffusion. As cations enter the brush region, they are almost
immediately bound to the brush and ion pairing equilibrium is quickly achieved
at the interface. This allows electrophoresis and diffusion to drive cations
further into the brush each time immediately binding and reaching ion pairing
equilibrium. This process slows until steady state is achieved around $\hat{t}=383$,
at which time ion pairing has reached equilibrium throughout the whole brush region
and any cation or anion movement due to diffusion, electrophoresis or the Born
force cancel out for zero net movement of ions.

\begin{figure*}
  \includegraphics[scale=0.40]{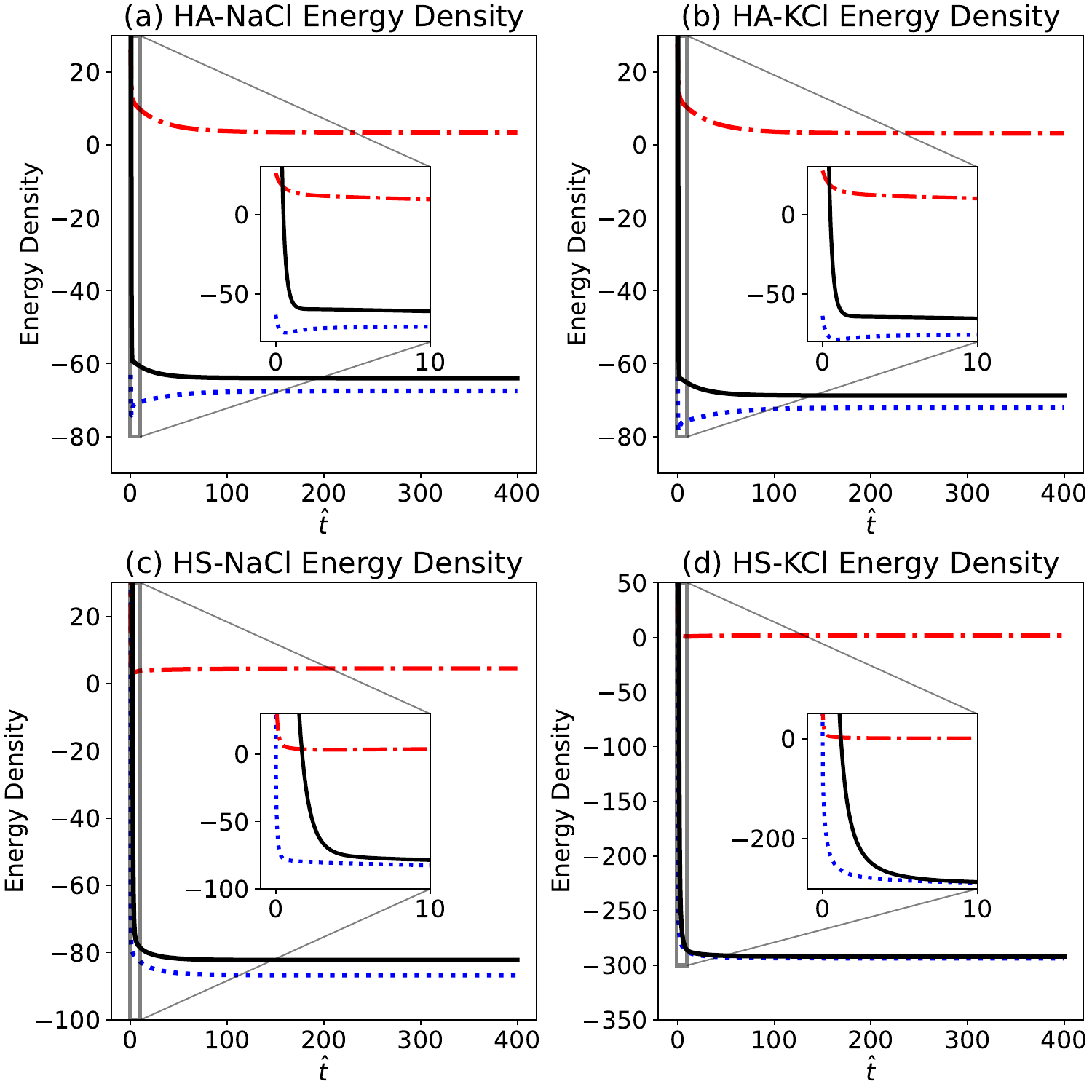}
  \caption{Dimensionless energy density curves versus time for the systems
  initially far from equilibrium. Total, $\hat{\mathcal{F}}$,
  (solid black curve), entropic/binding, $\hat{\mathcal{F}}_{3}$, (dotted blue curve), and Born
  solvation, $\hat{\mathcal{F}}_{4}$, (dashed dot red curve) energy densities versus time.
  Electrostatic field distribution, $\hat{\mathcal{F}}_{1}$, and electrostatic potential,
  $\hat{\mathcal{F}}_{2}$, energy densities quickly approach 0 and are not shown.}
  \label{energy-overlay}
\end{figure*}

We computed the dimensionless energy density at each time step. The results are
shown in Figure~\ref{energy-overlay} for each of the four systems. The total
energy density quickly decreases initially, and then slowly continues to
decrease to steady state. Looking at the individual energy terms, the electrostatic
field distribution and electrostatic potential energy densities reach a steady
state value of $0$ after quickly increasing and decreasing, respectively, thus
having no contribution to the steady state total energy density.

The entropic/binding energy density is the largest contributor to the steady state
total energy density. For the HS systems, the entropic/binding energy
density is dominated by the dissociation constant in the bound cation/brush term because it
is much less than $1$.  For all four systems,
the unbound brush term contributes significantly less to the entropic/binding
energy density. This can be explained qualitatively by looking at the
solid black curves in Figures~\ref{HANaCl_trans}--\ref{HSKCl_trans}. For all
four systems, the majority of the brush is bound to the cation in the steady state,
thus, there are fewer unbound brush ions to contribute to the entropic energy
density compared to the unbound cation, anion and bound cation/brush terms.

The Born solvation energy density has a small contribution to the steady state
total energy density. In the brush region at steady state, the unbound anion is
almost completely absent, and the majority of the cation is bound to the brush.
In the salt region, the value of the dimensionless permittivity, $\varepsilon_{1}$,
approaches $\varepsilon_{S}$. Thus, the unbound cation and anion terms have only
a small contribution to the Born solvation energy density.

\subsection{Physically realistic initial conditions}\label{init_real}
\begin{figure*}
  \includegraphics[scale=0.40]{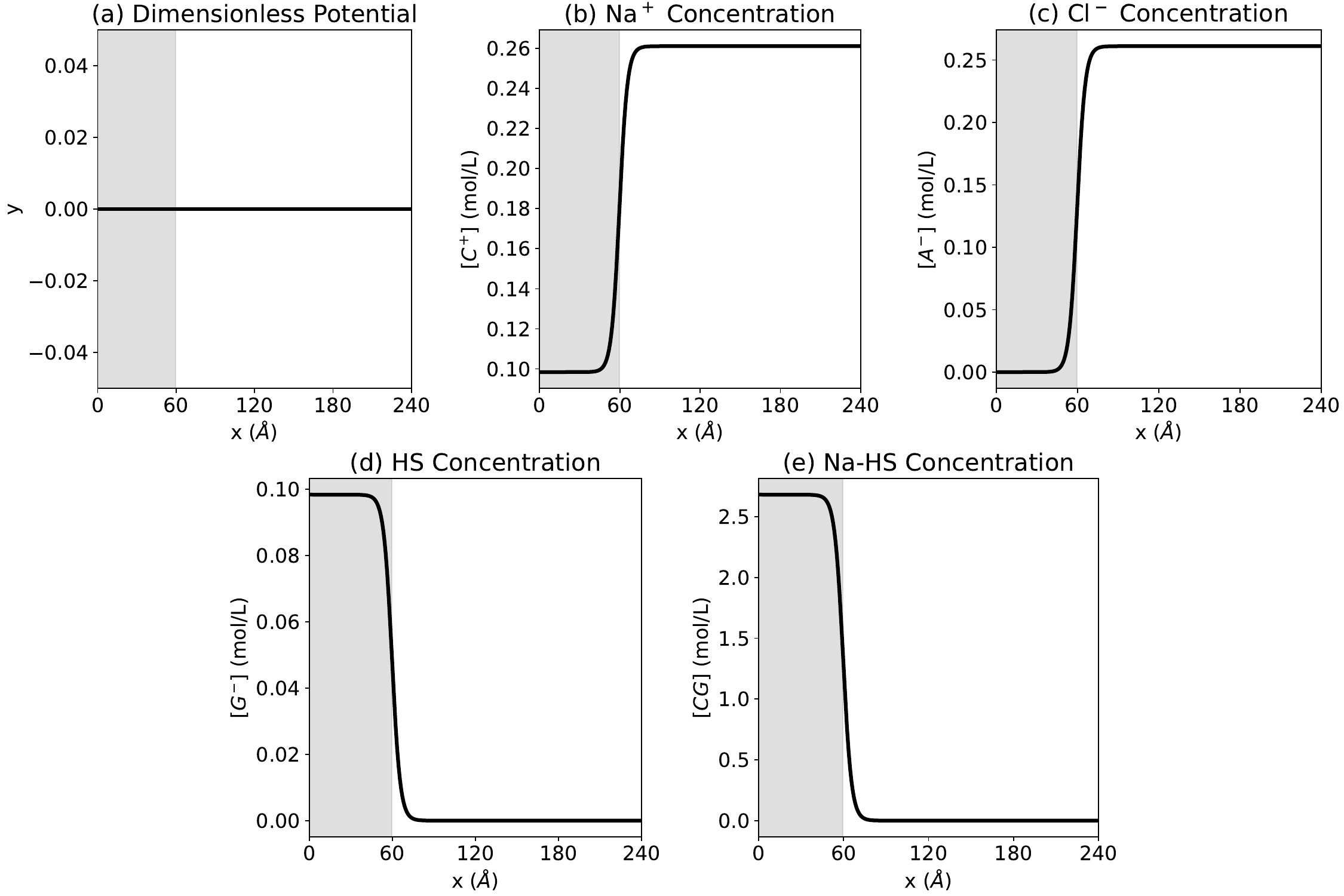}
  \caption{Physically realistic initial conditions for the HS-NaCl
  system. The shaded region represents the brush region.}
  \label{HSNaCl_phys_init_cond}
\end{figure*}

Now that we have shown that our model is capable of capturing the complex interactions
between diffusion, electrophoresis, the Born force, and ion pairing, we study the
same systems again using the parameters in Table~\ref{input_params}, but with
initial conditions that represent more physically realistic scenarios. Because
all four systems approach the same steady state solutions as before, we only
discuss the results for the HS-NaCl system.
The initial conditions are shown in Figure~\ref{HSNaCl_phys_init_cond} and start
the system in a state of electroneutrality. This was achieved by treating the brush
and salt regions independently. In the brush region, the cation and brush are
assumed to have already reached a state of equilibrium, with no anion present.
In the salt region, the cation and anion are also in a state of equilibrium.

\begin{figure*}
  \includegraphics[scale=0.40]{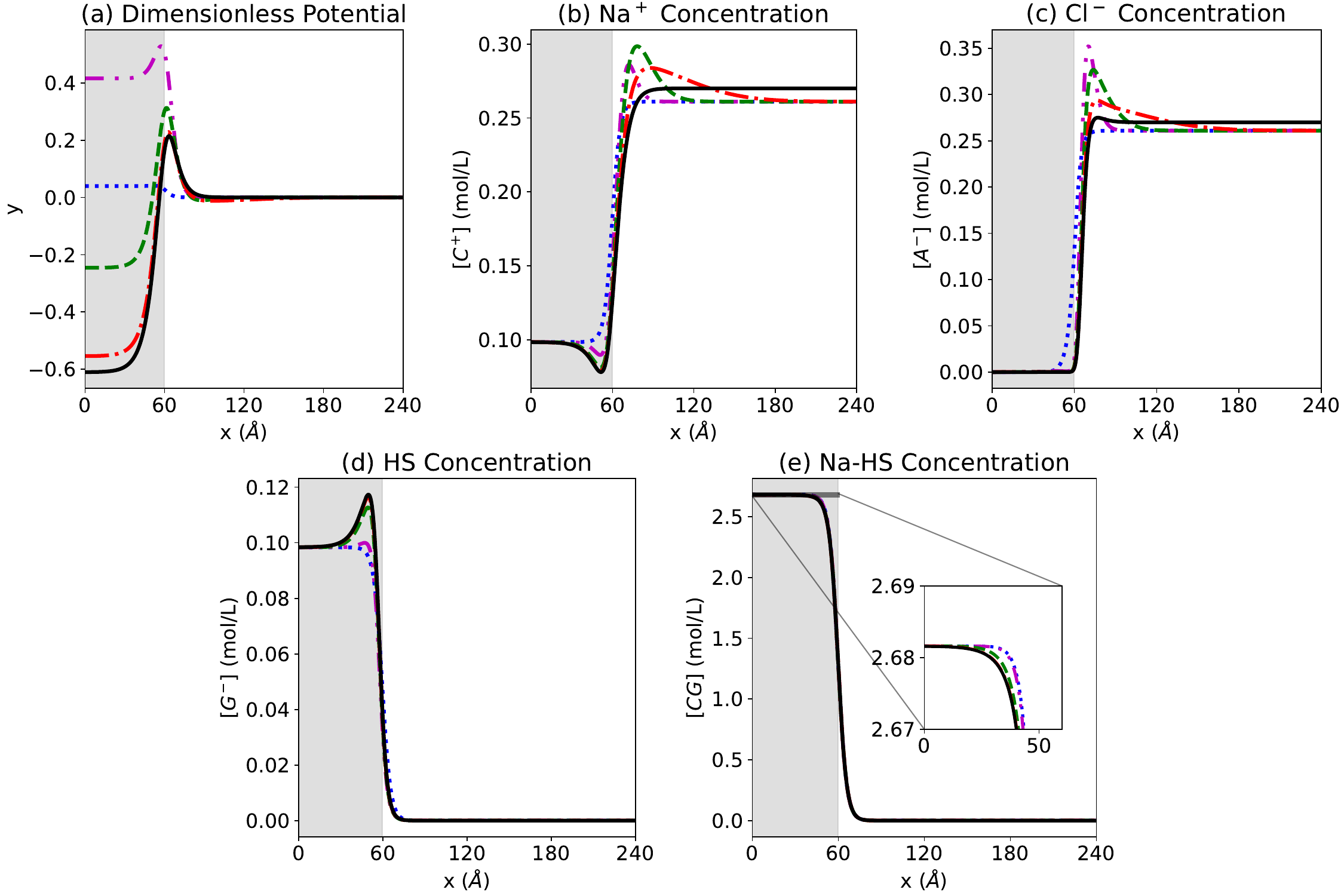}
  \caption{Transient results at $\hat{t}=0.01$ (dotted blue curve), $\hat{t}=0.51$
  (dashed dot dot magenta curve), $\hat{t}=2.5$ (dashed green curve), $\hat{t}=15$
  (dashed dot red curve), and $\hat{t}=322$ (solid black curve) for the HS-NaCl
  system with physically realistic initial conditions. The shaded
  region represents the brush region.}
  \label{HSNaCl_phys_trans}
\end{figure*}

At $\hat{t}=0$, the two regions begin to interact. The transient results for
selected time steps are shown in Figure~\ref{HSNaCl_phys_trans}. Unlike \S \ref{init_far},
at the initial time, the electric potential is zero everywhere, and thus there
is no electrophoresis. Diffusion wants to move both cations and anions into the
brush region, whereas the Born force wants to move/keep both out of the brush region.
Ion pairing is in equilibrium, except just to the left of the interface where
some net binding is occurring (this is an artifact of the value of $\beta$ chosen
in Table~\ref{input_params} used as a smoothing factor). The net impact for the
cation is a decrease in concentration to the left of the interface (Born and
binding dominate) and a slight increase to the right of the interface (Born and 
diffusion mostly cancel, however Born barely wins). For the anion, the Born force
is much stronger than diffusion and there is a decrease to the left of the interface
and an increase to the right of the interface (this is also an artifact of the
value of $\beta$). 

As time progresses, the electric potential in the brush initially increases,
reaching a maximum around $\hat{t}=0.3$, and then decreases towards steady state.
The electric potential to the right of the interface is higher than at either the
brush or salt boundaries. Thus, electrophoresis drives cations away from this
interface and into both the brush and salt regions and drives anions towards this
interface. At the same time, diffusion works in unison with, while the Born force
works against electrophoresis at the interface. The strength of the Born force
ultimately prevents any cations from entering the brush region and pushes all
anions out of the brush region. Diffusion then drives these ions towards their
steady state concentrations. Due to the decrease in cations to the left of the
interface, ion pairing switches to net unbinding leading to an increase in unbound
brush concentration and a decrease in bound cation-brush concentration to the
left of the interface until ion pairing equilibrium is achieved. Steady state
is achieved around $\hat{t}=322$, which is only approximately 60 ns faster compared
to starting with initial condition far from equilibrium.

\subsection{Two cation partitioning}\label{two_cations}
\begin{figure*}
  \includegraphics[scale=0.40]{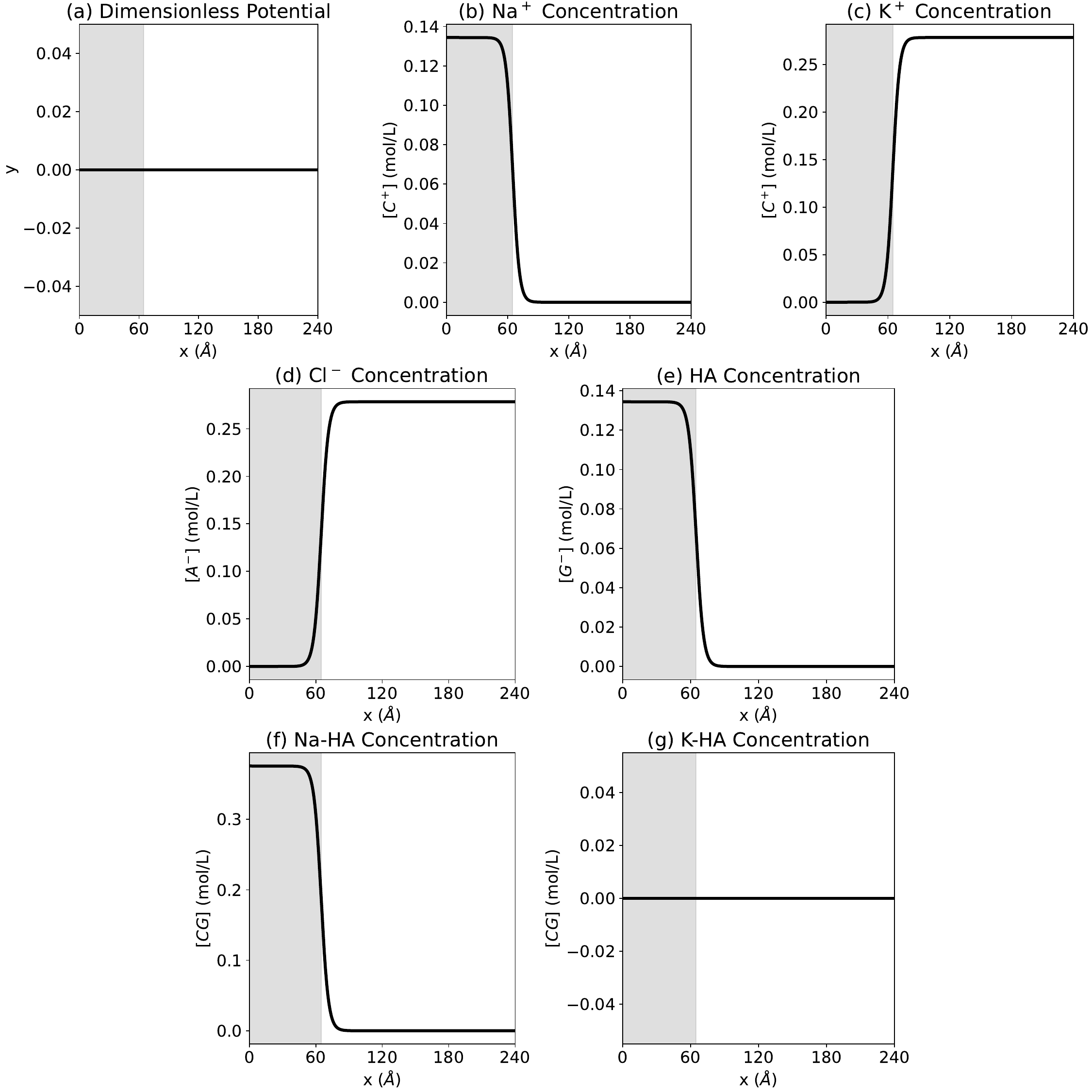}
  \caption{Initial conditions for the HA-Na/KCl system. The shaded
  region represents the brush region.}
  \label{HANa_KCl_init_cond}
\end{figure*}

Model equations (\ref{mpnp_model})-(\ref{mpnp_boundary}) can be easily expanded
to incorporate a second cation using the information provided in \S \ref{deriv-model}
and are not separately derived here. Including a second cation allows us to study
how binding reactions impact the transfer of cations. Using the parameters in
Table~\ref{input_params2} and the initial conditions in Figure~\ref{HANa_KCl_init_cond},
we start the system with Na$^{+}$ (cation 1) in equilibrium with the brush and
K$^{+}$ (cation 2) in equilibrium with Cl$^{-}$ (anion). Thus, Na$^{+}$ ions are
initially only in the brush region, and K$^{+}$ ions are initially only in the
salt region.

\begin{table}
  \caption{Input Parameters to two cation MPNP Model.}
  \label{input_params2}
  \begin{ruledtabular}
  \begin{tabular}{lllll}
    \hline
    \textbf{Parameter} & \textbf{HA-} \\
     & \textbf{Na/KCl} \\
    \hline
    $\hat{\ell}$ & $7.838$ \\
    $\hat{L}$ & $29.007$ \\
    $\varepsilon_{S}$ & $0.773$ \\
    $\varepsilon_{G}$ & $0.649$ \\
    $\beta$ & 0.1 \\
    $\hat{u}$ & 0.417 \\
    $\hat{r}_{c1}$ & 0.196 \\
    $d_{c1}$ & 1.167 \\
    $\alpha_{c1}$ & 1 \\
    $\hat{r}_{c2}$ & 0.236 \\
    $d_{c2}$ & 1.167 \\
    $\alpha_{c2}$ & 1 \\
    $\hat{r}_{a}$ & 0.0273 \\
    $d_{a}$ & 0.833 \\
    $\alpha_{a}$ & 1 \\
    $\hat{k}_{+1}$ & 1 \\
    $\hat{k}_{-1}$ & 0.172 \\
    $\hat{k}_{+2}$ & 1 \\
    $\hat{k}_{-2}$ & 0.114 \\
    \hline
  \end{tabular}
  \end{ruledtabular}
\end{table}

\begin{figure*}
  \includegraphics[scale=0.40]{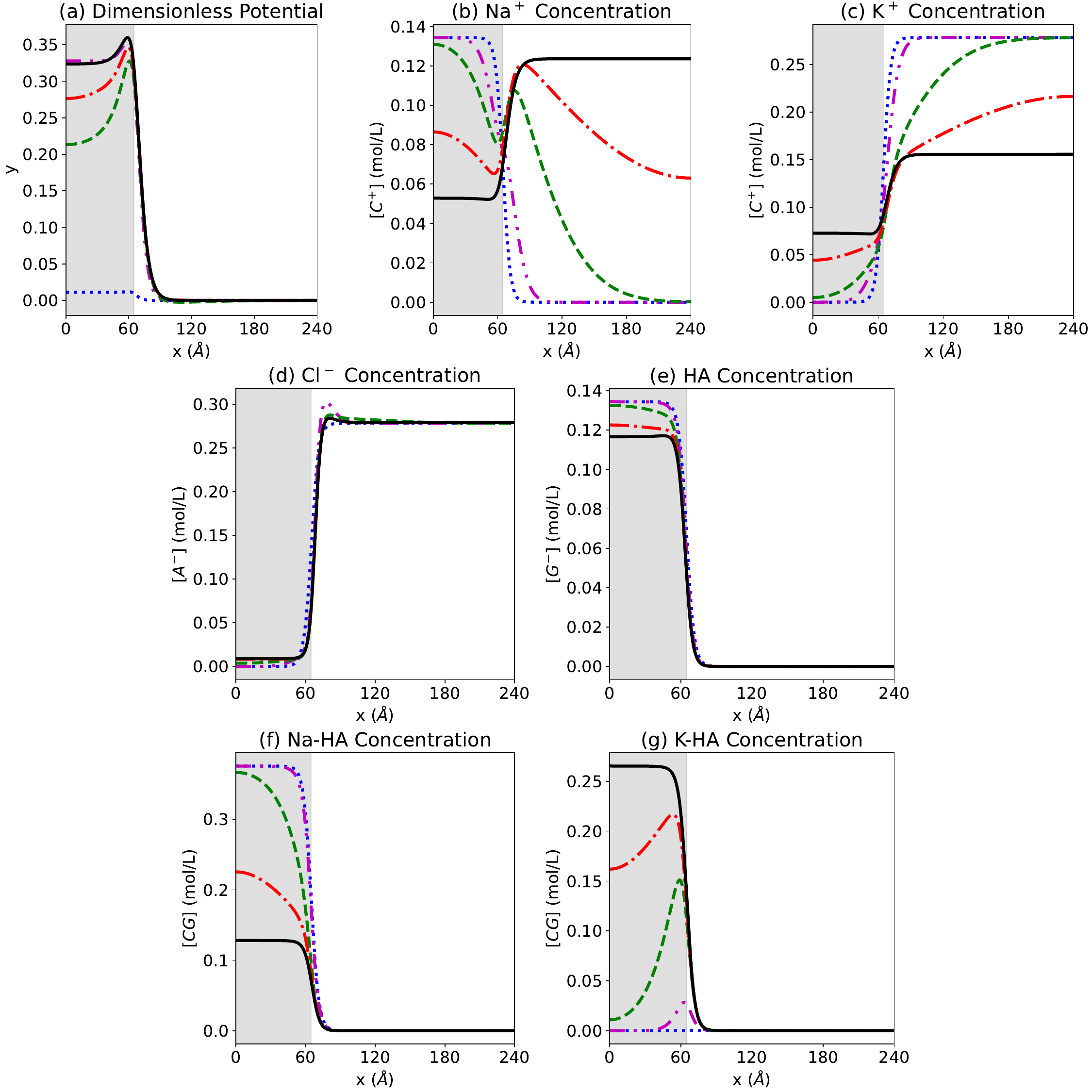}
  \caption{Transient results at $\hat{t}=0.01$ (dotted blue curve), $\hat{t}=1.12$
  (dashed dot dot magenta curve), $\hat{t}=18$ (dashed green curve), $\hat{t}=131$
  (dashed dot red curve), and $\hat{t}=963$ (solid black curve) for the HA-Na/KCl
  system. The shaded region represents the brush region.}
  \label{HANa_KCl_trans}
\end{figure*}

The transient results for selected time steps are captured in Figure~\ref{HANa_KCl_trans}.
At initial time, both regions are electroneutral and the electric potential is
zero everywhere yielding no electrophoresis. Na$^{+}$ ions want to diffuse from
the brush region into the salt region, whereas K$^{+}$ and Cl$^{-}$ ions want to
diffuse from the salt region into the brush region. The Born force wants to move
Na$^{+}$ ions from the brush region into the salt region and keep K$^{+}$ and
Cl$^{-}$ ions out of the brush region. Ion pairing is in equilibrium for Na$^{+}$,
while K$^{+}$ wants to begin binding with the brush. The net effect at the initial
time step is a decrease in Na$^{+}$ concentration and an increase in K$^{+}$ and
Cl$^{-}$ concentrations to the left of the interface. At the interface, Cl$^{-}$
concentration decreases. To the right of the interface, there is an increase in
Na$^{+}$ and Cl$^{-}$ concentrations and a decrease in K$^{+}$ concentration.

As time progresses, the electric potential in the brush increases and electrophoresis
begins to drive Na$^{+}$ out of the brush region, drive/keep K$^{+}$ out of the
brush region, and drive Cl$^{-}$ into the brush region. Diffusion continues to drive
K$^{+}$ into the brush region, overcoming both electrophoresis and the Born force.
Diffusion works together with electrophoresis for Cl$^{-}$, but the Born force
ultimately keeps most of the anion out of the brush. At the same time, Na$^{+}$
begins to unbind from the brush and is replaced with K$^{+}$, starting at the
interface and eventually throughout the brush. Most of the K$^{+}$ that enters
into the brush region winds up binding to the brush. As Na$^{+}$ moves out of the
brush region and into the salt region due to the combined efforts of diffusion,
electrophoresis, and the Born force, this results in a higher concentration to the
right of the interface. This leads to a complex diffusion process, which wants
to decrease Na$^{+}$ concentration in the brush region, increase the concentration
to the left of the interface, decrease the concentration to the right of the
interface, and finally, increase the concentration in the salt region.

Through the process of exchanging Na$^{+}$ for K$^{+}$ in the brush region, the
electric potential increases until $\hat{t}=1.12$, where it achieves a peak slightly
higher than the Donnan potential. The potential then begins a period of decreasing
until $\hat{t}=18$ and then increases toward steady state. Steady state is achieved
around $\hat{t}=963$, which is approximately $2.5-3$ times longer than the single
cation scenarios.

\section{Conclusion}\label{conclusion}
We have developed a time-dependent model and corresponding energy functional for
GAG brush/salt systems that builds upon previous PNP models
\cite{chen1992constant, liu2018analysis} incorporating Born solvation
\cite{liu2017incorporating} and binding energies. We compared the steady state
solution of our current model to our previous MPB model \cite{ceely2023mathematical}
using parameters derived from molecular simulation data \cite{sterling2021ion}
and found excellent agreement. We studied the difference in transient responses
using initial conditions close to and far from equilibrium and show that both
approach the same steady state. We also studied how binding reactions impact the
transfer of cations when one cation is initially in equilibrium with the GAG
brush and another cation is initially in equilibrium with the anion in the salt
region. Future work is to incorporate concentration dependent permittivity
\cite{ben2011dielectric, zhao2013influence} and GAG brush swelling
\cite{ehtiati2022specific} via Flory-Huggins \cite{zhang2021interfacial} and/or
other means \cite{zheng2023charge} to improve the model.

\bibliography{ref} 

\end{document}